\documentclass[twocolumn,aps,showpacs,prl,floatfix]{revtex4}
\usepackage{amsmath,amssymb,eucal,graphicx,hyperref}
\usepackage{dcolumn}% Align table columns on decimal point
\usepackage{bm}% bold math
\usepackage{array}
\usepackage{float}
\usepackage{supertabular}
\usepackage{longtable}
\usepackage{txfonts}
\begin{document}

\title{Evolutionary dynamics on degree-heterogeneous graphs}
\author{T. Antal}  \author{S. Redner} \author{V. Sood} \affiliation{Center of Polymer
  Studies and Department of Physics, Boston University, Boston,
  Massachusetts, 02215 USA}

\begin{abstract}
  
  The evolution of two species with different fitness is investigated on
  degree-heterogeneous graphs. The population evolves either by one
  individual dying and being replaced by the offspring of a random neighbor
  (voter model (VM) dynamics) or by an individual giving birth to an
  offspring that takes over a random neighbor node (invasion process (IP)
  dynamics).  The fixation probability for one species to take over a
  population of $N$ individuals depends crucially on the dynamics and on the
  local environment.  Starting with a single fitter mutant at a node of
  degree $k$, the fixation probability is proportional to $k$ for VM dynamics
  and to $1/k$ for IP dynamics.

\end{abstract}

\maketitle

In this letter, we investigate the likelihood for fitter mutants to
overspread an otherwise uniform population on heterogeneous graphs by
evolutionary dynamics.  Such a process underlies epidemic propagation
\cite{epidemics}, emergence of fads \cite{fads}, social cooperation
\cite{coop}, or invasion of an ecological niche by a new species
\cite{moran::MP,invasion,lieberman::EDG}.  At each update event, two
individuals from a total population $N$ are chosen at random.  One individual
replicates while the other dies and is replaced by the newly-born offspring,
so that $N$ remains constant.  A selective advantage, or fitness, also exists
in which each each individual may be a unit-fitness genotype ${\bf 1}$ or
genotype ${\bf 0}$ with lower fitness $1-s$, with $0<s<1$.  These fitnesses
determine the replication or death rates of each individual.  This selective
advantage leads to a dynamical competition in which selection dominates for
large populations, while random genetic drift \cite{kimura::NTE,ewens::MPG}
occurs for small populations or weak selection.

\begin{figure}[ht]
 \vspace*{0.cm}
\includegraphics*[width=0.35\textwidth]{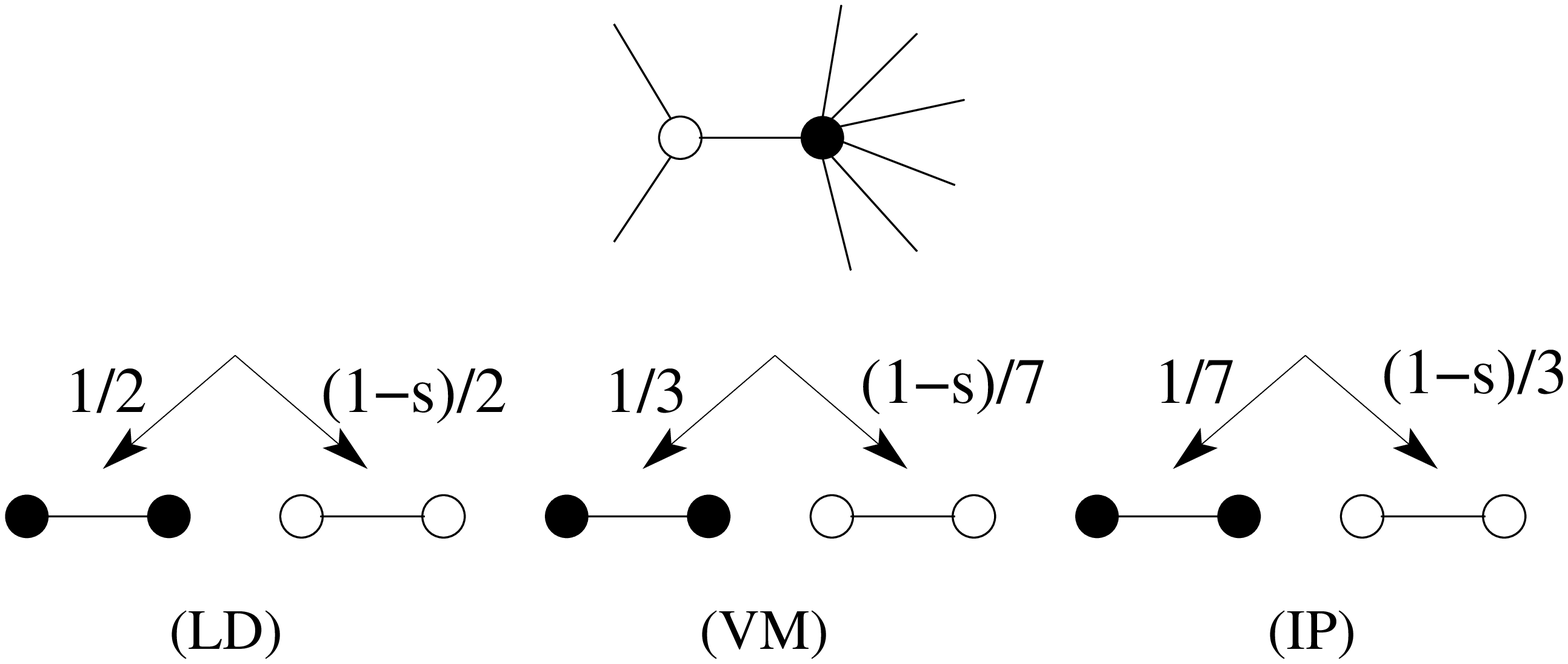}
\caption{Update illustration for two specific nodes.  Genotypes {\bf 0} and
  {\bf 1} are denoted by $\medcirc$ and $\medbullet$ respectively. Shown are the
  possible transitions and their respective relative rates due to the
  interaction of two nodes across a link for LD, VM, and IP dynamics.}
 \label{update}
\end{figure}

We consider three evolutionary models, distinguished by the order in which a
pair individuals replicate and die:
 
\noindent{\tt Biased Link Dynamics (LD):} A link is selected at random.  If
the individuals at the link ends are different, one of them is designated as
the ``donor'' with probability proportional to its fitness.  The replicate of
the donor then replaces the other individual: ${\bf 10} \to {\bf 11}$ with
probability $1/2$ while ${\bf 10}\to {\bf 00}$ with probability (1-s)/2
(Fig.~\ref{update}).

\noindent{\tt Biased Voter Model (VM):} An individual dies with probability
inversely proportional to its fitness, and is then replaced by the offspring
of a randomly-chosen neighbor.  Equivalently, death occurs randomly and
replacement is proportional to the fitness of the donor.  We implement the VM
by updating a randomly-chosen genotype {\bf 0} with probability 1, while the
fitter genotype {\bf 1} is updated with a probability $1-s$.  Each individual
in this death-first/birth-second process can equivalently be viewed as a
voter that adopts the opinion of a randomly-selected neighbor
\cite{liggett,krapivsky::VM,sood::VM}.

\noindent{\tt Biased Invasion Process (IP):} 
In this birth-first/death-second process, a randomly-chosen individual
replicates with probability proportional to its fitness, and its offspring
then replaces an individual at a randomly-chosen neighboring node
\cite{ewens::MPG,moran::MP,invasion}.

One genotype ultimately replacing all other genotypes in the population is
termed fixation.  An important, and easily checked fact is that these
evolutionary models are equivalent on degree-regular graphs; moreover, as we
will show, the fixation probability in LD can be obtained exactly,
independent of the underlying graph.  However, essential differences arise on
degree-heterogeneous networks \cite{sood::VM,SEM,C05} that may lead to an
enhancement of the fixation probability, as discovered previously for the IP
\cite{lieberman::EDG}.  Here we cast LD, VM, and IP on degree-heterogeneous
graphs within the same unifying framework to understand the interplay between
selection and random drift on the fixation probability.  By this approach, we
show that on degree-heterogeneous graphs the best strategy to reach fixation
with VM dynamics is for the fitter genotype to be on high-degree nodes.
Conversely, for IP dynamics, it is best for the fitter genotype to be on
low-degree nodes.

% LD, IP, and VM dynamics are equivalent on degree-regular graphs \cite{SEM};
% moreover the fixation probability in LD can be obtained exactly, independent
% of the underlying graph.  On degree-heterogeneous networks, however, the
% three dynamics lead to different outcomes.  For the IP, in particular,
% special graphs were identified that gave an enhanced fixation probability
% compared to that on regular graphs \cite{lieberman::EDG}.  In a similar vein,
% we will show that fixation for biased VM dynamics has a rich behavior that
% arises from the competition between selection and random drift.

We first study the evolution in VM dynamics.  We symbolically represent the
state of the system by $\eta$.  In an elemental time interval $\delta t$ we
choose a random node $x$.  If the genotype at this node at time $t$, denoted
as $\eta^t(x)$, equals ${\bf 0}$, then node $x$ is updated by choosing a
random neighbor $y$ and setting $\eta^{(t+\delta t)}(x)=\eta^t(y)$
(Fig.~\ref{update}).  However if $\eta^t(x) = {\bf 1}$, the VM update is
implemented with probability $1-s$. This update rule can be written as
\begin{eqnarray}
\label{trans::voter}
{\bf P}[\eta\!\to\!\eta_x]=
\sum_y \frac{A_{x y}}{Nk_x}\Big\{[1\!-\!\eta(x)]\eta(y)
+(1\!-\!s)\eta(x)[1\!-\!\eta(y)]\Big\}\!,
\end{eqnarray}
% \begin{eqnarray}
% \label{trans::voter}
% {\bf P}[\eta\to\eta_x]&=&\frac{1}{N}
% \sum_y \frac{A_{x y}}{k_x}\Big\{[1-\eta(x)]\eta(y)\nonumber \\
% &~&~~~~~~~~~+(1-s)\eta(x)[1-\eta(y)]\Big\}\,,
% \end{eqnarray}
where $\eta_x$ denotes the state obtained from $\eta$ by changing only the
genotype at node $x$.  The first term describes the update step for the case
where $(\eta(x),\eta(y))= ({\bf 0},{\bf 1})$ and $x, y$ are connected.
Each of the nearest neighbors $y$ of $x$ may be selected with probability
$A_{xy}/k_x$.  Here $A_{xy}$ is the adjacency matrix whose elements equal 1
if $xy$ are connected and zero otherwise.  The second term in
Eq.~\eqref{trans::voter} is explained analogously.

For degree-heterogeneous graphs, the density $\rho_k$ of genotype {\bf 1} at
nodes of degree $k$ increases by $1/N_k$ with probability ${\bf F}_k(\eta)$ and
decreases by $1/N_k$ with probability ${\bf B}_k(\eta)$ in an elemental update,
where
\begin{equation}
\label{FBkdef}
\begin{split}
{\bf F}_k(\eta) &=  \frac{1}{k N}\mathop{{\sum}'}_{x y~} A_{x y} [1-\eta(x)]\eta(y)\\
{\bf B}_k(\eta) &= \frac{1-s}{k N}\mathop{{\sum}'}_{x y~}A_{x y} \eta(x)[1-\eta(y)]
\end{split}
\end{equation}
are the forward ({\bf 0} $\to$ {\bf 1}) and backward ({\bf 1} $\to$ {\bf 0})
evolution rates.  The primes on the sums denote the restriction that the
degree of nodes $x$ equals $k$.  The sum over all $k$ then gives the total
transition rate of Eq.~\eqref{trans::voter}.  We seek the fixation
probability $\Phi$ to the state consisting entirely of genotype {\bf 1} as a
function of the initial densities of {\bf 1}.  This probability obeys the
backward Kolmogorov equation $G\Phi=0$ \cite{karlin}, subject to the boundary
conditions $\Phi(0)=0$ and $\Phi(1)=1$.  In the diffusion approximation, the
generator $G$ of this equation may be expressed as a sum of the changes in
$\rho_k$ over all $k$,
\begin{equation}
\label{gendef}
G = \frac{1}{\delta t}\sum_k 
\left[ \delta\rho_k ({\bf F}_k\!-\!{\bf B}_k)\partial_k \!+\! \frac{(\delta\rho_k)^2}{2}({\bf F}_k
\!+\!{\bf B}_k)\partial_k^2\right]\,,
\end{equation}
with $\delta\rho_k = 1/N_k = 1/(N\, n_k)$ the change in $\rho_k$ in a single
update of a node of degree $k$, and
$\partial_k\equiv\frac{\partial}{\partial\rho_k}$.

For the special case of degree-regular graphs, where $k_x=k$ for all nodes,
both sums in Eq.~\eqref{FBkdef} count the total number $\alpha$ of active
links between different genotypes
\begin{equation}
 \alpha = \frac{1}{N \mu_1} \sum_{x,y} A_{xy} \eta(x) [1-\eta(y)] \,,
\end{equation}
where the moments of the degree distribution are defined by $\mu_n \equiv
N^{-1} \sum_x k_x^n = \sum_k k^n n_k$.  The generator thus reduces to
\begin{equation}
\label{diffusion::regular}
G  = \alpha\left[s\partial_{\rho} + \frac{1}{N}(1-\frac{s}{2})\partial_{\rho}^2\right]\,,
\end{equation}
where we use $\delta\rho = \delta t = 1/N$.  In this form, the convection and
diffusion terms differ by a factor $\mathcal{O}(s N)$.  Thus selection
dominates when the population $N$ is larger than $\mathcal{O}(1/s)$, while
random genetic drift is important otherwise.  

Notice that the probability of increasing the density of genotype {\bf 1} at
each update is a factor $1/(1-s)$ larger than the probability of decreasing
the density.  By its construction, this same bias arises for LD on general
networks.  As a consequence of this bias, the evolutionary process underlying
fixation is the same as the absorption of a uniformly biased random walk in a
finite interval \cite{karlin}, from which the fixation probability is
\begin{equation}
\label{fixation::regular}
\Phi(\rho) = \frac{1-(1-s)^{N\rho}}{1-(1-s)^{N}}\to \frac{1-e^{-sN\rho
    /(1-s/2)}}{1-e^{-sN/(1-s/2)}}\,.
\end{equation}
The former is the exact discrete solution of $G\Phi=0$ on a finite network,
while the latter continuum limit represents the solution to $G\Phi=0$ in the
diffusion approximation.  These results apply for all three models on
degree-regular graphs and for LD on general graphs.

For degree-heterogeneous graphs, the conserved quantity for neutral dynamics
($s=0$) is the average degree-weighted density $\omega_1$
\cite{sood::VM,SEM}, where the degree-weighted moments are
\begin{equation}
  \omega_{n} = \frac{1}{N\mu_{n}} \sum_x k_x^n \eta(x) = \frac{1}{\mu_{n}} \sum_k k^n n_k \rho_k\,,
\end{equation}
while the overall density $\rho$ of genotype {\bf 1} is no longer conserved.
The existence of this new conservation law suggests that we study the time
evolution of the expectation value of $\omega_1$ which we henceforth denote
as $\omega$ for notational simplicity.  Since $\omega(\eta_x) = \omega(\eta)
+ k_x(1-2\eta(x))/\mu_1 N$,
\begin{eqnarray}
\label{evolution::omega::general}
\partial_t\omega &=& \frac{1}{\delta t}\sum_x
\left[\omega(\eta_x) - \omega(\eta)\right]{\bf P}[\eta\to\eta_x]\nonumber\\
&=& \frac{s}{\mu_1 N}\sum_{x,y}A_{x y} \eta(x)(1-\eta(y)) = s \alpha\,.
\end{eqnarray}
Notice that $\omega$ is conserved in the absence of selection ($s=0$) a
feature that ultimately stems from the update rate being inversely
proportional to node degree [Eq.~\eqref{trans::voter}].  To evaluate the
expression in Eq.~\eqref{evolution::omega::general} we make the mean-field
assumption that the degrees of connected nodes in the graph are uncorrelated.
Thus we replace the elements of the adjacency matrix by their expected
values, $A_{x y} = k_x k_y/\mu_1 N$.  This assumption simplifies
Eq.~(\ref{evolution::omega::general}) to $\partial_t\omega =
s\omega(1-\omega)$, with solution $\omega(t)^{-1} =
1-[1-\omega(0)^{-1}]e^{-st}$.

\begin{figure}[ht]
 \vspace*{0.cm}
\includegraphics*[width=0.405\textwidth]{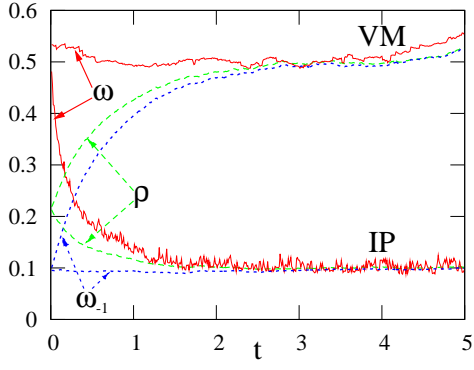}
\caption{Moments of the {\bf 1} density in the biased VM and biased IP on a
  network of $10^4$ nodes with a power-law degree distribution $n_k \sim
  k^{-\nu}$ ($\nu=2.5$), and no correlations between node degrees.  Nodes
  with degree larger than the mean degree are initialized to {\bf 1} while
  all other nodes are {\bf 0}.  For the VM, $s=8.5 \times 10^{-4}$, while for
  the IP, $s=10^{-4}$.  }
\label{traj}
\end{figure}

For uncorrelated graphs, Eqs.~\eqref{FBkdef} simplify to
\begin{eqnarray}
\label{update::MR}
{\bf F}_k(\eta) = n_k \omega (1-\rho_k), \quad {\bf B}_k(\eta) = (1-s) n_k (1-\omega) \rho_k\,.
\end{eqnarray}
Thus the time evolution of the expectation value of $\rho_k$ is
\begin{equation}
\label{evolution::rhok::MR}
\partial_t\rho_k = \frac{\delta\rho_k({\bf F}_k - {\bf B}_k)}{\delta t}
= \omega-\rho_k + s(1-\omega)\rho_k\,.
\end{equation}
To solve this equation we combine it with $\partial_t\omega =
s\omega(1-\omega)$ to give $\partial_t(\omega-\rho_k) =
-(\omega-\rho_k)(1-s(1-\omega))$, with solution
\begin{equation}
\label{evolution::rhok::MR::solution}
\omega(t)\!-\!\rho_k(t) = e^{-t}[\omega(0)\!-\!\rho_k(0)]\{\omega(0) + [1\!-\!\omega(0)]e^{-s t}\}\,.
\end{equation}
For small selective advantage ($s\ll 1$), this equation involves two
distinct time scales.  On a time scale of order one, all the $\rho_k$
become equal to $\omega$, whereas the evolution of $\omega$ occurs on a
longer time scale of order $s^{-1}\gg 1$ (Fig.~\ref{traj}).

We now determine the fixation probability simply by replacing the $\rho_k$ by
$\omega$ in the forward and backward rates {\bf F} and {\bf B} in
Eqs.~\eqref{update::MR}.  In a similar vein, we replace the derivative
$\partial_k$ by $(k n_k/\mu_1)\partial_{\omega}$ \cite{sood::VM}.  Then the
generator in Eq.~\eqref{gendef} becomes
\begin{eqnarray}
\label{Kolmogorov::MR}
G &=& s\sum_k\left(\frac{k n_k}{\mu_1}\right)\omega(1-\omega)\partial_{\omega}\nonumber\\
&+& \frac{1}{N}\left(1-\frac{s}{2}\right)\sum_k\left(\frac{k^2 n_k}{\mu_1^2}\right)
\omega(1-\omega)\partial^2_{\omega}\nonumber\\
&=& \omega(1-\omega)\left[ s\partial_{\omega} + 
\frac{\mu_2}{\mu_1^2 N}\left(1-\frac{s}{2}\right)\partial^2_{\omega}\right]\,.
\end{eqnarray}
Apart from an overall constant for the time scale, this generator is
identical to that of degree-regular graphs \eqref{diffusion::regular} if we
replace $N$ by an effective population size $N_{\rm eff} \equiv
N\mu_1^2/\mu_2$.  For a network of $N$ nodes with a power-law degree
distribution, $n_k\propto k^{-\nu}$, $N_{\rm eff}$ scales as \cite{sood::VM}
\begin{equation}
\label{Neff}
N_{\rm eff} \equiv \frac{\mu_1^2 N}{\mu_2}\sim
\begin{cases}
N  & \nu>3\,;\cr
N^{(2\nu-4)/(\nu-1)} & 2<\nu<3\,;\cr
\mathcal{O}(1) & \nu<2\,,
\end{cases}
\end{equation}
with logarithmic corrections for $\nu=2$ and $\nu=3$.  Thus $N_{\rm eff}$
becomes much less than $N$ when $\mu_2$ diverges; this occurs when $\nu<3$.
A similar change in the effective size of the population is observed for
biological species evolving in a spatially heterogeneous environment
\cite{moran::MP,barton::1997}.

\begin{figure}[ht]
\vspace*{0.cm}
\includegraphics*[width=0.405\textwidth]{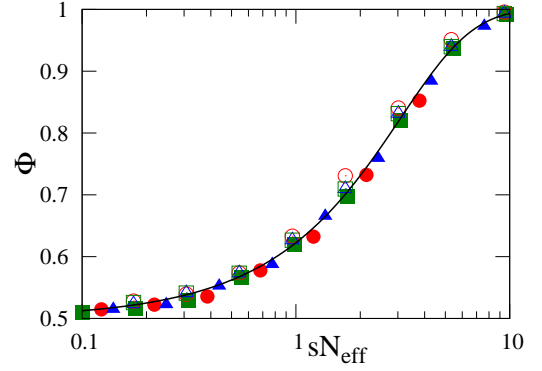}
\caption{Scaling plot of fixation probabilities for VM (filled) and IP
  dynamics (open symbols).  
%Here $N_{\rm eff} =\mu_1^2 N/\mu_2$ for the VM
%  and $N_{\rm eff}= N$ for IP.  
  Data are for degree-uncorrelated graphs with $N=10^3$, $\mu_1 = 8$, and
  degree distribution exponent $\nu = 2.1$ ($\medcirc$), $ 2.5$
  ($\triangle$), or $3.0$ ($\square$).  Initially each node is a mutant with
  probability 1/2, ($\omega_1=\omega_{-1}=1/2$).  The curve corresponds to
  Eqs.~\eqref{fixation::voter} or \eqref{fixation::invasion}.}
\label{surv-scaled}
\end{figure}

\begin{figure}[ht]
\vspace*{0.cm}
\centerline{\includegraphics*[width=0.45\textwidth]{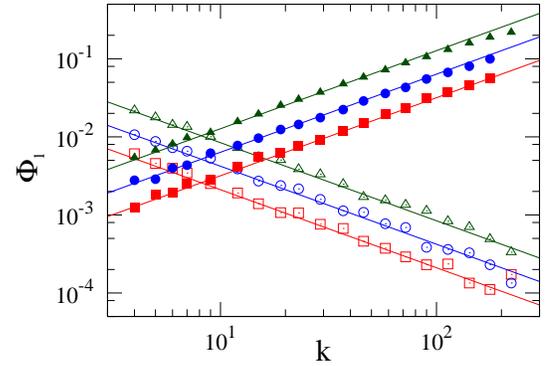}}
\caption{Fixation probability of a single mutant initially at a node of
  degree $k$ on an uncorrelated power-law degree distributed ($n_k \sim
  k^{-\nu}, ~\nu=2.5$) graph with $N=10^3$ and $\mu_1 = 8$.  The empty
  symbols correspond to IP dynamics with $s = 0.004$ ($\square$), $s = 0.008$
  ($\medcirc$) and $s = 0.016$ ($\triangle$); the filled symbols correspond
  to VM dynamics with $s = 0.01$, ($\blacksquare$), $s = 0.02$ ($\medbullet$)
  and $s=0.08$ ($\blacktriangle$).  The solid lines, with slopes $+1$ and
  $-1$, correspond to the second of Eqs.~\eqref{fixation::voter-k} and
  \eqref{fixation::invasion-k}.}
\label{surv-one}
\end{figure}

The solution to  $G\Phi=0$, with $G$ given by Eq.~\eqref{Kolmogorov::MR} is
\begin{equation}
\label{fixation::voter}
\Phi(\omega) = \frac{1-e^{-sN_{\rm eff}\omega /(1-s/2)\,}}{1-e^{-sN_{\rm eff}/(1-s/2)}}\,.
\end{equation}
Our numerical data for the fixation probability shows both excellent scaling
and agreement with this functional form for $\Phi$ (Fig.~\ref{surv-scaled}).
Eq.~\eqref{fixation::voter} also provides the fixation
probability when the system starts with a single mutant at a node of degree
$k$:
\begin{equation}
\label{fixation::voter-k}
\Phi_1 =
\begin{cases}
k/(N\mu_{1}) & s\ll 1/N_{\rm eff}\,; \\
sk\mu_1/\mu_2 & 1/N_{\rm eff} \ll s\ll 1\,.
\end{cases}
\end{equation}
The crucial feature is that the fixation probability of a single fitter
mutant is proportional to the degree of the node that it initially occupies
(Fig.~\ref{surv-one}).  Notice also that because the relative effect of
selection versus random genetic drift is determined by the variable
combination $sN_{\rm eff}$, random genetic drift can be important for much
larger populations compared to the case of degree-regular graphs.  In fact,
for a power-law graph with $\nu<2$, random genetic drift prevails for all
population sizes.

We now study fixation in the complementary biased invasion process.  Here a
randomly-selected individual reproduces with probability proportional to its
fitness; hence the transition probability is
\begin{eqnarray}
\label{trans::invasion}
{\bf P}[\eta\!\to\!\eta_x]\!=\!
\sum_y\! \frac{A_{x y}}{Nk_y}\Big\{[1\!-\!\eta(x)]\eta(y)
\!+\!(1\!-\!s)\eta(x)[1\!-\!\eta(y)]\Big\},
\end{eqnarray}
Notice an essential difference between VM and IP dynamics.  In the VM the
transition rate is proportional to the inverse degree $k_x$ of the node of
the disappearing genotype [Eq.~\eqref{trans::voter}], while in the IP the
transition rate is proportional to the inverse degree $k_y$ of the node of
the reproducing genotype [Eq.~\eqref{trans::invasion}].

For degree-uncorrelated graphs, the transition probabilities are
\begin{equation}
\label{update::MR-IP}
 {\bf F}_k(\eta) = \frac{k}{\mu_1} n_k \rho (1-\rho_k)\,, 
\quad {\bf B}_k(\eta) = \frac{k(1-s)}{\mu_1} n_k (1-\rho) \rho_k\,.
\end{equation}
Consequently the time evolution of $\rho_k$ is given by, in analogy with
Eq.~\eqref{evolution::rhok::MR},
$\partial_t \rho_k = \frac{k}{\mu_1}[\rho-\rho_k + s \rho_k(1-\rho)]$,
%\begin{equation}
%\label{rhok-dot}
%\partial_t \rho_k = \frac{k}{\mu_1}[\rho-\rho_k + s \rho_k(1-\rho)]\,,
%\end{equation}
from which low-order moments obey the equations of motion:
\begin{eqnarray*}
\partial_t \omega_{-1} &=& \frac{s}{\mu_1 \mu_{-1}} \rho(1-\rho) \,,\\
\partial_t \rho &=& \rho - \omega + s \omega(1-\rho)\,, \\
\partial_t \omega &=& \frac{\mu_2}{\mu_1}[\rho-\omega_2 + s \omega_2(1-\rho)]\,.
\end{eqnarray*}
In contrast to the VM, the conserved quantity in the unbiased IP is
$\omega_{-1}$, the inverse degree-weighted frequency.  For the biased IP,
$\omega_{-1}$ becomes the most slowly changing quantity (see
Fig.~\ref{traj}).  Hence we transform all derivatives with respect to
$\rho_k$ in the generator to derivatives with respect to $\omega_{-1}$ to
yield
\begin{equation}
\label{diffusion::invasion}
G = \frac{\omega_{-1}(1-\omega_{-1})}{\mu_1 \mu_{-1}} \left[ 
s\frac{\partial}{\partial \omega_{-1}} + \frac{1}{N}(1-\frac{s}{2})\frac{\partial^2}{\partial \omega_{-1}^2}\right]\,,
\end{equation}
from which, in close analogy with our previous analysis of the VM, the
fixation probability is
\begin{equation}
\label{fixation::invasion}
\Phi(\omega_{-1}) = \frac{1-e^{-sN\omega_{-1}/(1-s/2)}}{1-e^{-sN/(1-s/2)}}.
\end{equation}

From Eq.~\eqref{fixation::invasion}, the effective population size $N_{\rm
  eff}$ equals $N$, contrary to VM dynamics (Eq.~\eqref{fixation::voter}).
More strikingly, the fixation probability of a single mutant acquires the
non-trivial dependence of the degree $k$ of the occupied node
(Fig.~\ref{surv-one})
\begin{equation}
\label{fixation::invasion-k}
\Phi_1= 
\begin{cases}
1/(Nk\mu_{-1}) & s\ll 1/N \,; \\
s/(k\mu_{-1}) & 1/N \ll s \ll 1\,.
\end{cases}
\end{equation}

To conclude, mutants are more likely to fixate in the voter model (VM) when
they are initially on high-degree nodes [Eq.~\eqref{fixation::voter-k}],
while in the invasion process (IP) fixation is more probable when mutants
start on low-degree nodes [Eq.~\eqref{fixation::invasion-k}].  This behavior
is understandable simply.  In the VM, a well-connected individual is more
likely to be asked his opinion before he asks one of his neighbors.  In the
IP, a mutant on a high-degree node is more likely to be invaded by a neighbor
before the mutant itself can invade.  Thus network heterogeneity leads to
effective evolutionary heterogeneity.

We can also understand the evolution when a mutant appears at a random node
on a graph.  In the selection-dominated regime ($sN_{\rm eff} \gg 1$) of the
VM, we average Eq.~\eqref{fixation::voter-k} over all nodes and find that the
fixation probability on degree-uncorrelated graphs is smaller by a factor
$\mu_1^2/\mu_2\le1$ than that on regular graphs.  Thus a heterogeneous graph
is an inhospitable environment for a mutant that evolves by VM dynamics.
Conversely, performing the same average of Eq.~\eqref{fixation::invasion-k}
over all nodes, the fixation probability for the IP is the same on all
degree-uncorrelated graphs.  Finally, in the small-selection limit ($sN_{\rm
  e} \ll 1$), the node average fixation probability is the same for both the
VM and IP on degree-uncorrelated graphs.

\acknowledgments{We gratefully acknowledge financial support from the Swiss
  National Science Foundation under Grant No.\ 8220-067591NSF (TA) as well as
  US National Science Foundation Grant No.\ DMR0535503 (SR and VS).  }

\end{document}